\documentclass{article}

\usepackage{geometry}
\usepackage{graphicx}
\usepackage{amssymb}
\usepackage{amsfonts}
\usepackage{amsmath}
\usepackage{amsthm}
\usepackage{mathrsfs}
\usepackage{color}
\usepackage{natbib}
\usepackage{comment}
\usepackage{subfig}
\usepackage{enumitem}
\usepackage[colorlinks=true, linkcolor=blue, citecolor=blue, urlcolor=blue]{hyperref}
\usepackage[font=small,labelfont=bf,labelsep=period,width=12.9cm]{caption}

\title{RLE Plots:\ Visualising Unwanted Variation \\ in High Dimensional Data}
\author{Luke C. Gandolfo\footnote{Bioinformatics Division, The Walter and Eliza Hall Institute of Medical Research, Melbourne, Victoria, 3070, Australia.} \footnote{School of Mathematics and Statistics, University of Melbourne, Melbourne, Victoria, 3070, Australia.} \ and Terence P. Speed\footnotemark[1] \footnotemark[2]}
\date{}

\begin{document}

\maketitle
\begin{abstract}
Unwanted variation can be highly problematic and so its detection is often crucial. Relative log expression (RLE) plots are a powerful tool for visualising such variation in high dimensional data. We provide a detailed examination of these plots, with the aid of examples and simulation, explaining what they are and what they can reveal. RLE plots are particularly useful for assessing whether a procedure aimed at removing unwanted variation, i.e.\ a normalisation procedure, has been successful. These plots, while originally devised for gene expression data from microarrays, can also be used to reveal unwanted variation in many other kinds of high dimensional data, where such variation can be problematic.
\end{abstract}

\section{Introduction}

Relative log expression (RLE) plots are a simple, yet powerful, tool for visualising unwanted variation in high dimensional data. They were originally devised for analysing data from gene expression studies involving microarrays, e.g.\ see \citet{bolstad2005} and \citet{brettschneider2008}. Such studies generate high dimensional data: expression levels (i.e.\ activity levels) of many thousands of genes are measured simultaneously using a microarray (one for each studied individual). Studies often involve many individuals and therefore many microarrays. Unfortunately, the data generated is often affected by \emph{unwanted variation}, i.e.\ variation caused by technical factors and not by the biology of interest. There are many causes of such variation (see \citealp{scherer2009}). For example, batches of samples may be processed in different laboratories which operate at different temperatures leading to variation between the batches, i.e.\ a so-called \emph{batch effect}. Moreover, the temperature of a particular laboratory may be quite variable throughout the day leading to additional variation \emph{within} a batch. In any particular study, however, the physical causes of such variation will typically be unknown. This unwanted variation can be so large that comparing gene expression values between samples, often the main objective of such a study, can no longer be sensibly done; doing so can lead to false positives, false negatives, or both. Thus, it is crucially important to be able to detect the presence of unwanted variation. This is what RLE plots were devised to do.

Because of their ability to detect unwanted variation, RLE plots are particularly useful for assessing whether a \emph{normalisation} procedure, i.e.\ a procedure aimed at removing unwanted variation, has been successful: a ``bad" plot would suggest a failure to normalise (e.g.\ see \citealp{gagnon2012}). RLE plots, while originally devised for microarray data, can also be used to reveal unwanted variation in many other kinds of high dimensional data, e.g. metabolomic, proteomic, and RNA sequencing data, to name a few.

Our aim here is to provide a detailed examination of these plots, with the aid of examples and simulation. We begin by explaining what an RLE plot is, then we describe what it can reveal. We then discuss a number of important points to keep in mind when interpreting the plot. To make our discussion concrete we frame it in terms of one kind of data: microarray data. Nearly everything we say applies \emph{mutatis mutandis} to other kinds of high dimensional data.

\section{What is an RLE plot?}

We suppose that our microarray expression data (after log transformation) is organised into a matrix with $m$ rows, each representing a microarray sample, and $n$ columns, each representing the expression measurements for a particular gene across the samples. Let $y_{ij}$ be the log expression for gene $j$ in sample $i$, and let $y_{\ast j}$ denote the $j$th column of the matrix $[y_{ij}]$. An RLE plot is constructed as follows:
\begin{enumerate}
\item For each gene $j$, calculate its median expression across the $m$ samples, i.e.\ $\text{Med}(y_{\ast j})$, then calculate the deviations from this median, i.e.\ calculate $y_{ij} - \text{Med}(y_{\ast j})$, across the $i$s.
\item For each sample, generate a boxplot of all the deviations for that sample.
\end{enumerate}
We use the median, a robust measure of centre, to protect against outliers.

To furnish an example we consider data from a study by \citet{vawter2004}. The aim of this study was to find genes that are expressed differently between the brains of men and women. The study design is briefly summarised as follows (the full details need not concern us). Brain tissue samples were obtained (post-mortem) from 5 men and 5 women, with tissue taken from three distinct brain regions of each person, producing 30 tissue samples. Each of these samples were split into three portions, thereby producing three identical groups of 30 samples, and each group was sent to a different laboratory to be analysed with one of two types of microarray. Data for this study is available on Gene Expression Omnibus (GSE2164).

\begin{figure}
\begin{center}
\includegraphics[scale=0.3]{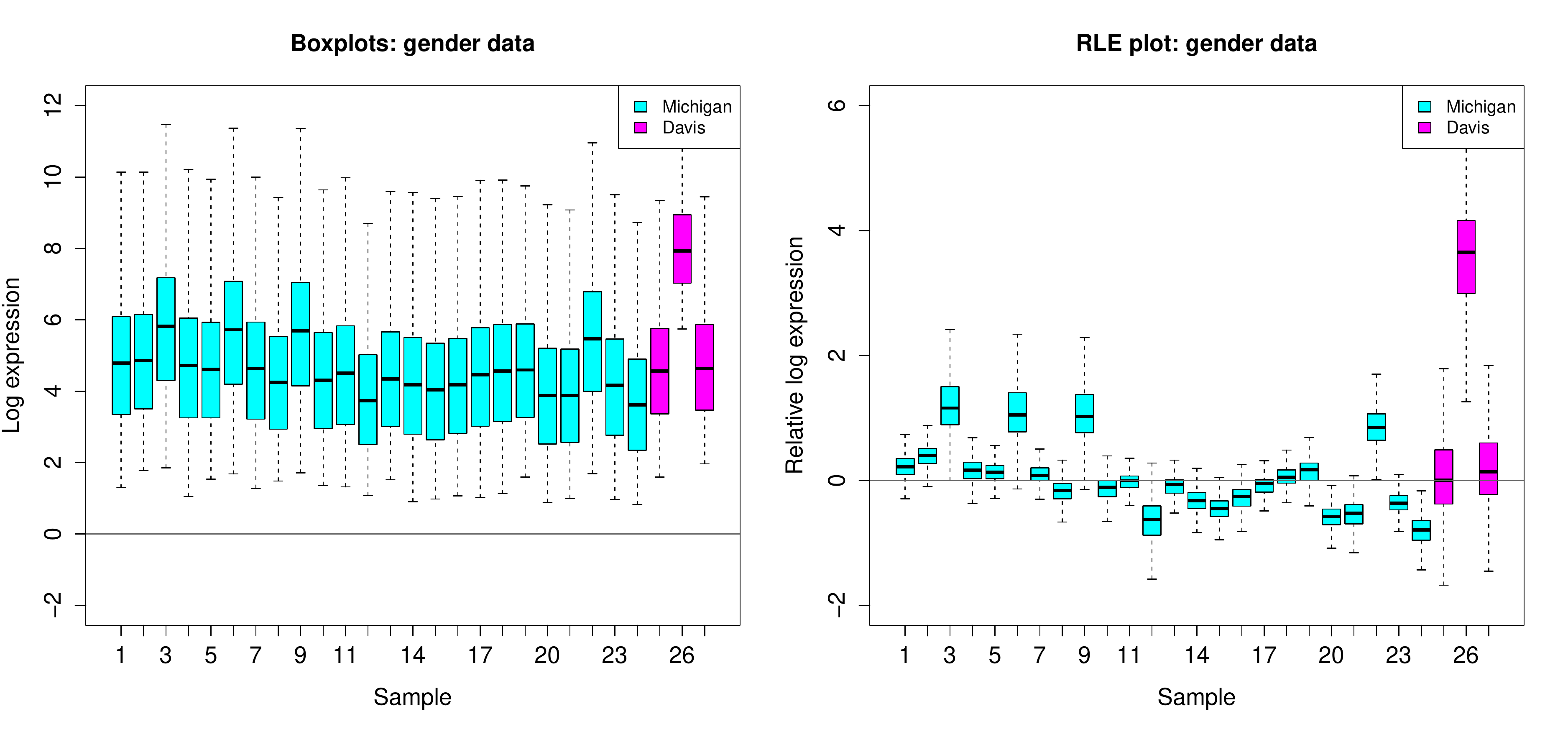}
\caption{\textbf{Example: gender data}. Boxplots and RLE plot of the gender data with colour coding for the University of Michigan and UC Davis laboratories.}
\label{fig: gender}
\end{center}
\end{figure}

For our purposes we restrict our attention to a small subset of 27 samples: these were all the samples analysed with the same kind of microarray (the Affymetrix HG-U95A microarray, measuring the expression of 12,626 genes), but processed at two different laboratories (24 at University of Michigan, 3 at UC Davis). We henceforth refer to this as the \emph{gender data}. We processed the data, performing background correction and summarisation (but not the default quantile normalisation), with the RMA package (\citealt{irizarry2003}), then generated an RLE plot using the steps described above (see Figure \ref{fig: gender}). For contrast, we also generated standard boxplots of the data by skipping the first of the above steps.

\section{What can RLE plots reveal?}

\subsection{RLE plots can reveal unwanted variation}

The most obvious feature an RLE plot reveals is sample heterogeneity. For example, the plot for the gender data shows large differences between samples. A deeper interpretation of an RLE plot can be made if we assume the following: 
\begin{itemize}[leftmargin=2cm,rightmargin=2cm]
\item[(A)]  Expression levels of a majority of genes are unaffected by the biological factors of interest. 
\end{itemize}
This is often a plausible assumption. In the gender study, for example, it is plausible to assume that only a minority of genes will be expressed differently between men and women for the same brain region. The same applies to different brain regions in the same individual, male or female.

In ideal circumstances, i.e.\ where no unwanted variation is present, under assumption (A) the log expression measurements for a majority of genes would simply consist of a mean plus a random variation about that mean: $y_{ij} = \mu_j + \epsilon_{ij}$, where $\epsilon_{ij}$ has zero mean and constant variance (depending on the gene $j$) across the samples. Since $\text{Med}(y_{\ast j}) \approx \mu_j$, by subtracting the median when constructing an RLE plot, we would obtain
\begin{equation*}
\label{ }
y_{ij} - \text{Med}(y_{\ast j}) = \mu_j + \epsilon_{ij} - \text{Med}(y_{\ast j}) \approx \mu_j + \epsilon_{ij} - \mu_j = \epsilon_{ij},
\end{equation*}
i.e.\ we remove the variation between genes, leaving only the variation between samples. So, in ideal circumstances, an RLE plot would only display $\epsilon_{ij}$s: the boxplots would be roughly centred on zero and would roughly be the same size. Thus, under (A), sample heterogeneity is a sign of unwanted variation.

The RLE plot for the gender data is far from the above ideal, and so reveals substantial unwanted variation. We see unwanted variation both between and within batches as indicated by the varying \emph{position} and \emph{widths} of the boxplots. Note that the substantial within batch variation, between the Michigan samples, is only dimly apparent from the standard boxplots, but is brought into sharp relief in the RLE plot. Also note that variation, in the form of varying boxplot widths, is \emph{only} revealed in the RLE plot. In the standard boxplots, this variation is obscured by the variation between genes; it is only revealed by removing the between gene variation in the construction of an RLE plot.

\subsection{Simulated data}

We have seen that ``bad" RLE plots reveal unwanted variation in two ways: varying boxplot position and varying boxplot width. What kinds of effects, in a statistical sense, produce these features in a plot? To help answer this question we use simulation. We simulate log expressions $y_{ij}$, for gene $j$ in sample $i$, under the following model:
\begin{equation}
\label{model}
y_{ij} = \mu_j + \theta_i + \gamma_{ij} + \epsilon_{ij},
\end{equation}
where $\mu_j$ and $\theta_i$ are \emph{additive} gene and sample effects, respectively, $\gamma_{ij}$ is a \emph{non-additive} effect, and $\epsilon_{ij}$ is a random error. We will use a simple \emph{multiplicative} form for the non-additive effect: 
\begin{equation*}
\label{ }
\gamma_{ij} = \lambda (\theta_i-\overline{\theta})(\mu_j-\overline{\mu}),
\end{equation*}
where $\overline{\theta}$ and $\overline{\mu}$ are the means over $i$ and $j$, respectively, and $\lambda$ is a constant. This form for $\gamma_{ij}$ is the kind discussed by \citet{tukey1949} and \citet{mandel1969, mandel1971}. Note that by subtracting row and column means in the product we obtain a non-additive effect which is ``orthogonal" to the additive effects. Also note that, while $\lambda$ can take any value, below we simply use it as an indicator to switch the non-additive effect ``on" or ``off". Using \eqref{model}, we simulate $y_{ij}$  as follows:
\begin{enumerate}
  \item[$\bullet$] For each gene $j$, simulate $\mu_j \sim N(m_\mu,s_\mu^2)$.
  \item[$\bullet$] For each sample $i$, simulate $\theta_i \sim N(m_\theta,s_\theta^2)$.
  \item[$\bullet$] For a fixed gene $j$, for each sample $i$ simulate $\epsilon_{ij} \sim N(0,\sigma_j^2)$, where we simulate $1/\sigma_j^2 \sim \text{Gamma}(\alpha, \beta)$.
\end{enumerate}
This model implies that mean expression varies across genes, and each gene varies differently across samples. We can obtain a batch effects by assigning different values of $m_\theta$ to different batches of samples. Note that we require $\mathbb{E}(\sigma_j^2) < s_\mu^2$, i.e.\ the variation across samples is typically less than the variation between genes, since this is usually a property of real microarray data. 

\begin{figure}[t!]
\begin{center}
\subfloat[]{\includegraphics[scale=0.29]{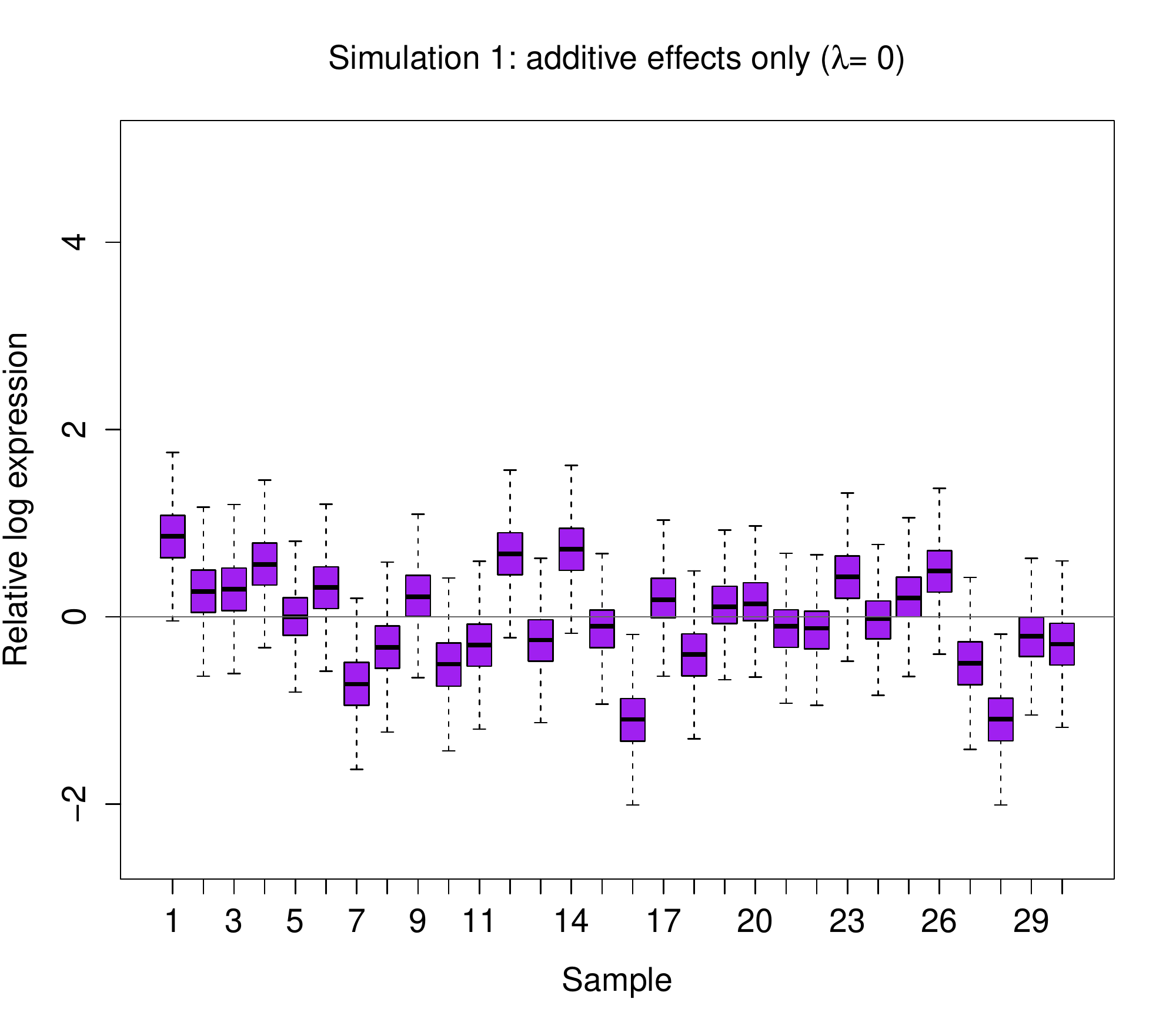}}
\subfloat[]{\includegraphics[scale=0.29]{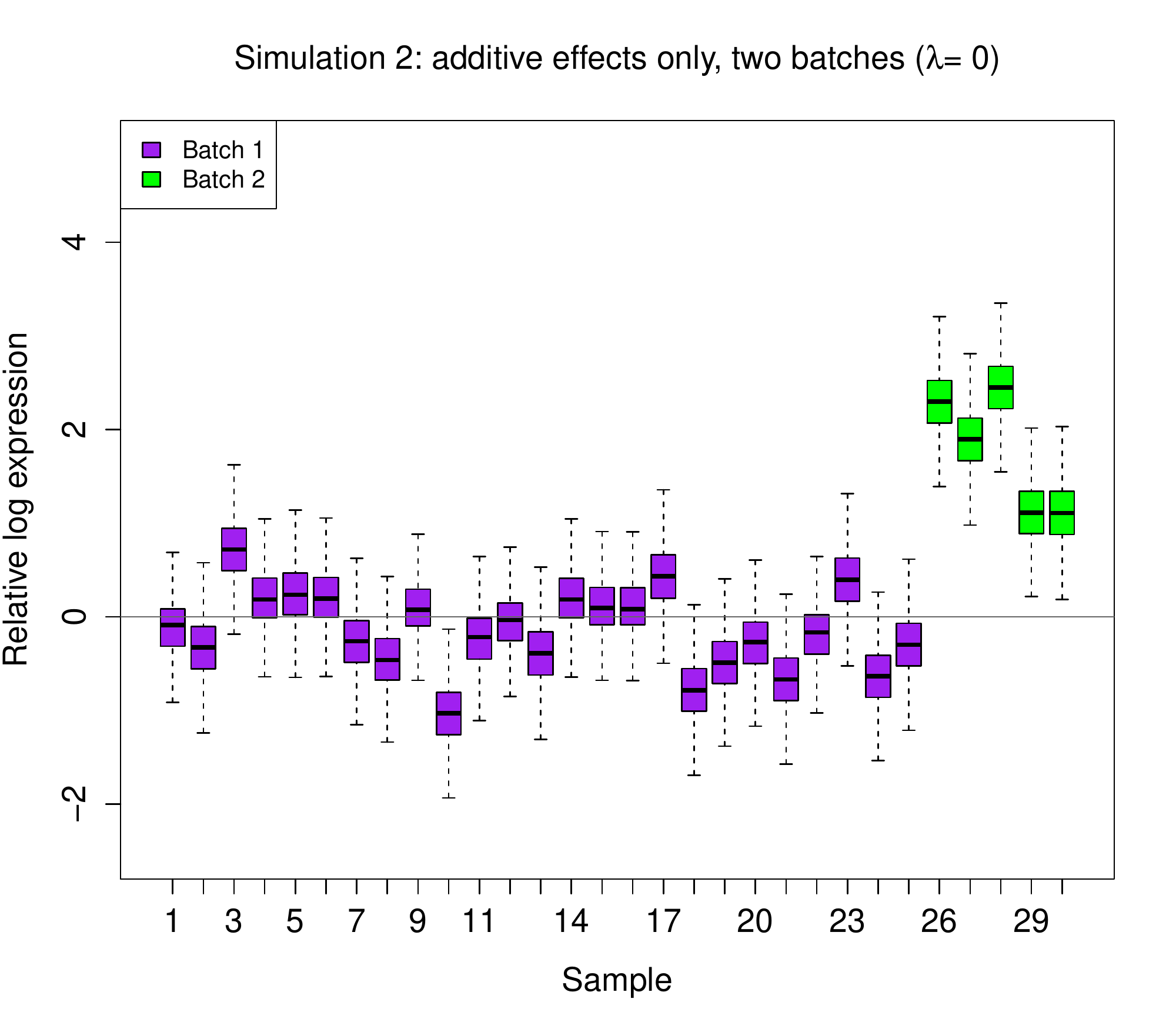}} \\
\subfloat[]{\includegraphics[scale=0.29]{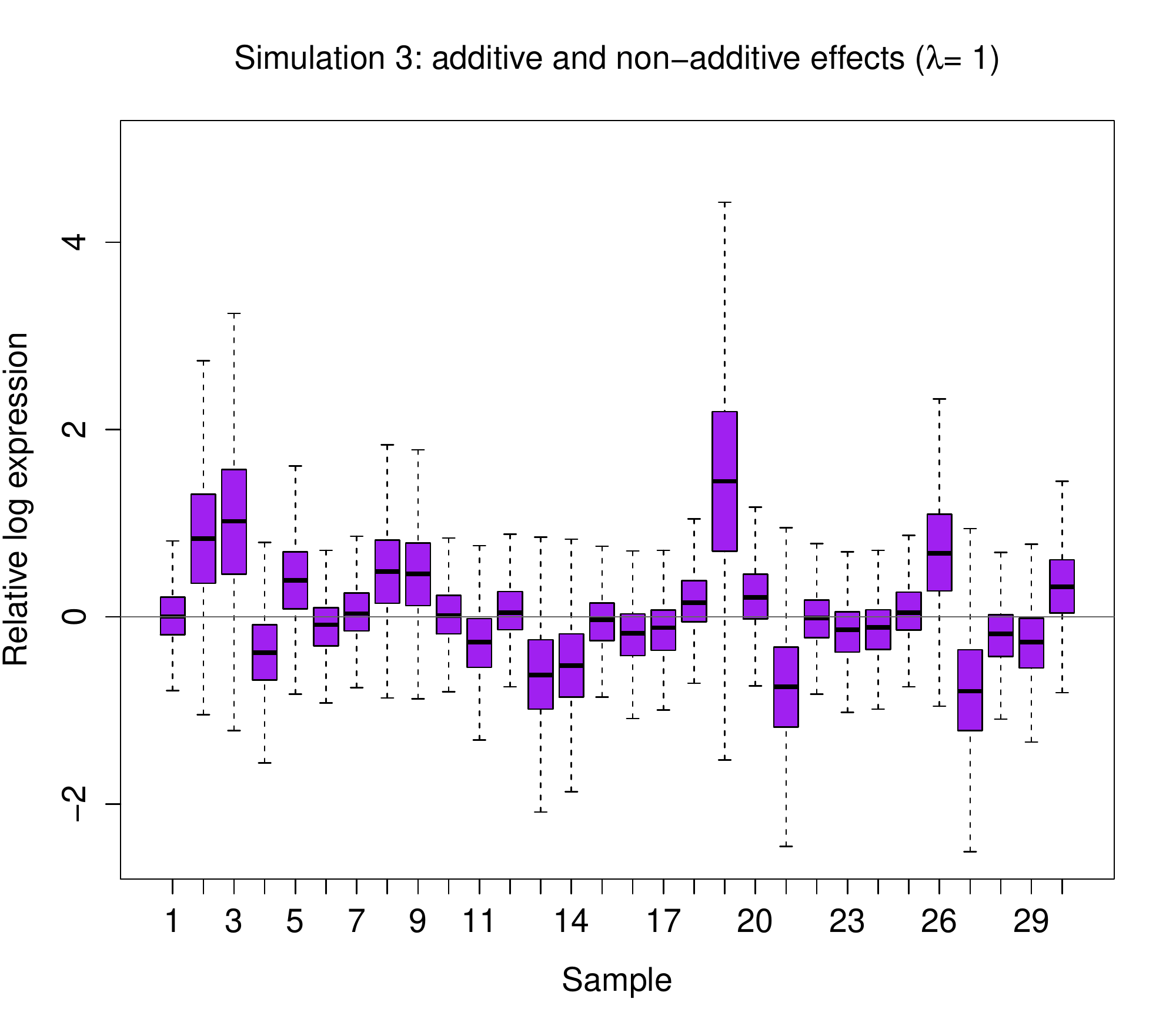}} 
\subfloat[]{\includegraphics[scale=0.29]{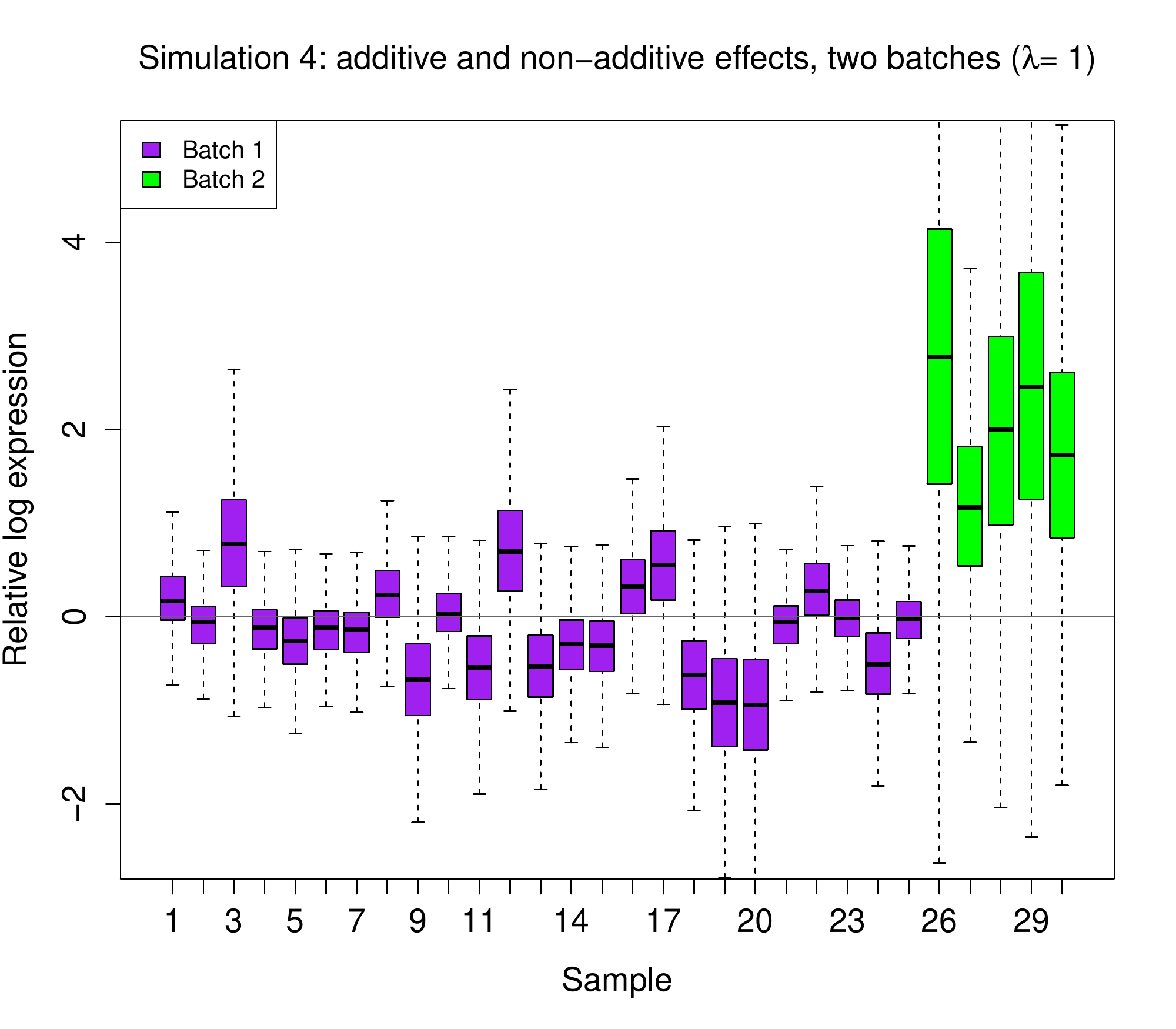}}
\caption{\textbf{Simulated data: RLE plots.} (a) additive effects only; (b) additive effects only, in two batches; (c) additive and non-additive effects; (d) additive and non-additive effects, in two batches.}
\label{fig: simulation}
\end{center}
\end{figure}

With this model, we simulate four different data sets each with $m = 30$ samples and $n = 10,000$ genes:
\begin{enumerate}
  \item \emph{Additive effects only:} $m_\theta = 0$ and $\lambda = 0$ for all $i$.

  \item \emph{Additive effects only, in two batches:} $m_\theta = 0$ for $i \in [1,25]$, $m_\theta = 2$ for $i \in [26,30]$, and $\lambda = 0$ for all $i$.
  
  \item \emph{Additive and non-additive effects:} $m_\theta = 0$ and $\lambda = 1$ for all $i$.
    
  \item \emph{Additive and non-additive effects, in two batches:} $m_\theta = 0$ for $i \in [1,25]$, $m_\theta = 2$ for $i \in [26,30]$, and $\lambda = 1$ for all $i$.
\end{enumerate}
In all instances, we set $m_\mu = 5, s_\mu^2 = 0.5, s_\theta^2 = 0.5, \alpha = 10, \beta = 1$. The latter two parameters give $\mathbb{E}(\sigma_j^2) = 0.11 < s_\mu^2$, as required. We then generate RLE plots for each data set (see Figure \ref{fig: simulation}).

These results show, perhaps not surprisingly, that shifting boxplots are produced by additive sample effects. However, perhaps more surprisingly, we see that additive sample effects are not sufficient to produce variation in the boxplot widths; this feature only appeared when non-additive sample effects were present. We do not claim that these are the \emph{only} kinds of statistical effects that produce ``bad" RLE plots -- there are clearly others. These effects, however, seem to be some of the most important, as we will see next.

\subsection{Real data}

\begin{figure}[h!]
\begin{center}
\subfloat[]{\includegraphics[scale=0.26]{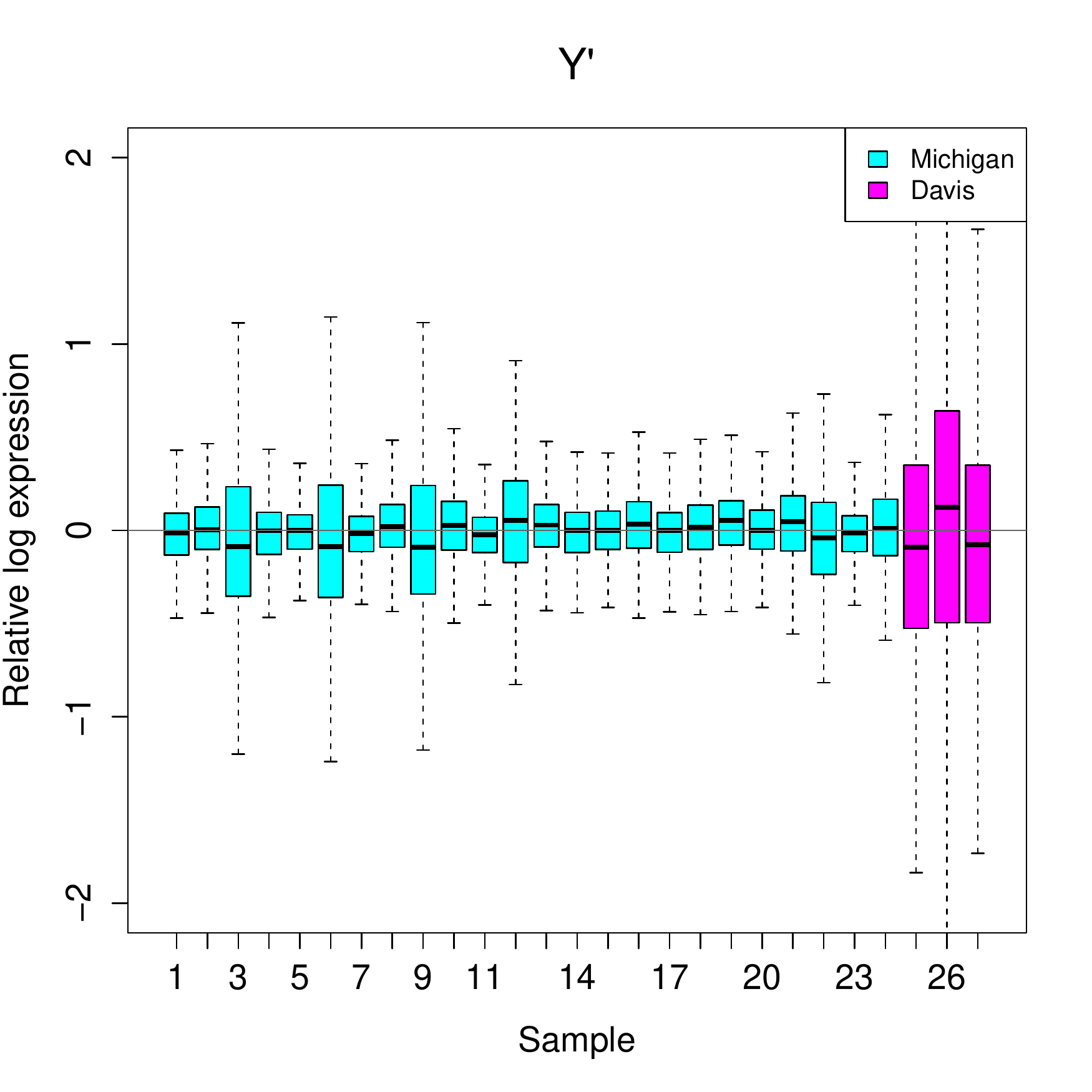}}\\
\subfloat[]{\includegraphics[scale=0.35]{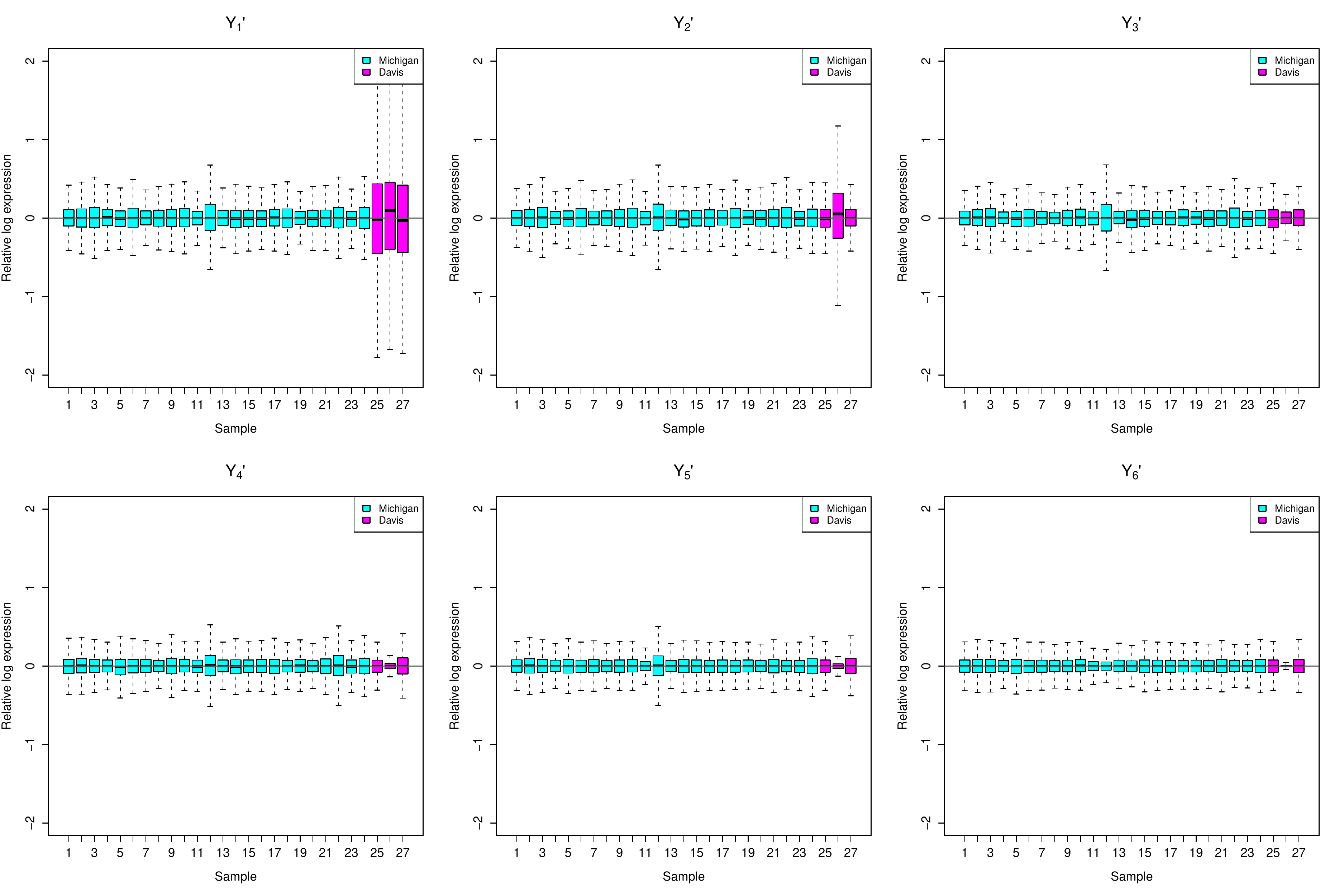}}
\caption{\textbf{Gender data: removing additive and non-additive sample effects}: (a) RLE plot of the gender data with the additive sample effect removed; (b) RLE plots of the gender data with the additive and successive non-additive sample effects removed, i.e.\ $\mathbf{Y}_p^\prime$ for $p = 1, \ldots, 6$}
\label{fig: gender svd}
\end{center}
\end{figure}

Can additive and non-additive sample effects help explain ``bad" RLE plots for \emph{real} data? The answer is ``yes", in many instances. Take, for example, the gender data again. Let $\mathbf{Y}$ be the matrix containing the gender data, where $y_{ij}$ is the log expression for gene $j$ in sample $i$. First, we try removing the additive sample effect by calculating $y^\prime_{ij} = y_{ij} - y_{\cdot j}$, where the dot indicates averaging over the subscript replaced by the dot. The RLE plot for $\mathbf{Y^\prime}$ (see Figure \ref{fig: gender svd}a) suggests that additive sample effects provide an explanation for much of the variation in boxplot position. 

To investigate the non-additive effects we examine the following residual, which is the standard estimate for the non-additive component of a linear model:
\begin{equation*}
\label{ }
d^\prime_{ij} = y_{ij} + y_{\cdot \cdot} - y_{i\cdot} - y_{\cdot j}.
\end{equation*}
Following \citet{mandel1969,mandel1971}, we can partition the non-additive component of the data by applying the singular value decomposition (SVD) to the matrix $\mathbf{D'}$, whose entries are these residuals. First observe that the SVD of $\mathbf{D'}$ can be written as a sum of (rank 1) matrices:
\begin{equation*}
\label{ }
\mathbf{D'} = \mathbf{U}\mathbf{\Sigma}\mathbf{V}^T = \sum_{k = 1}^r \sigma_k \mathbf{v}_k\mathbf{u}_k^{T} = \sum_{k = 1}^r \mathbf{M}_k,
\end{equation*}
where $r$ is the rank of $\mathbf{D'}$, $\sigma_k$ is the $k$th singular value, and $\mathbf{v}_k$ and $\mathbf{u}_k$ are $k$th left and right singular vectors, respectively. Now observe that the entries of these matrices have the form $[M_{ij}] = \sigma_k u_i v_j$, i.e.\ a product with one factor indexed by $i$, the other indexed by $j$, all scaled by $\sigma_k$. In other words, we have decomposed the non-additive component of the data into a sum of non-additive effects of simple multiplicative type. So, by subtracting $\mathbf{M}_k$ matrices from $\mathbf{Y^\prime}$ we can remove non-additive sample effects from the data, in addition to the additive sample effect removed previously. Given this, define $\mathbf{Y}_p^\prime$ as follows:
\begin{equation*}
\label{ }
\mathbf{Y}_p^\prime = \mathbf{Y^\prime} - \sum_{k = 1}^p \mathbf{M}_k,
\end{equation*}
where $p = 1, \ldots, r$. We generate RLE plots of $\mathbf{Y}_p^\prime$ for $p = 1, \ldots, 6$ (see Figure \ref{fig: gender svd}b).
We see that as $p$ increases the RLE plots approach their ideal appearance: the boxplots line up around zero and become roughly the same size. This suggests that non-additive sample effects provide the rest of the explanation for the variation in boxplot position and much of the explanation for the variation in boxplot widths. 

Since additive and non-additive sample effects are not the only kinds of statistical effects that can conceivably  produce ``bad" RLE plots, we do not wish to claim that these effects provide the only explanation for the features seen in the RLE plot for the gender data, only that they provide a \emph{possible} explanation.

\section{Discussion}

We commented in the introduction that RLE plots are particularly useful for assessing whether a normalisation procedure, i.e.\ a procedure that attempts to remove unwanted variation, has been successful; a ``bad" plot indicates a failure to normalise. It is important to note, however, that achieving an ideal RLE plot after applying a normalisation procedure does \emph{not} necessarily mean the procedure has succeeded. The procedure may have succeeded in removing the unwanted variation, but may have also removed the biological signal of interest, i.e.\ the differences in expression of a minority of genes. An RLE plot cannot diagnose whether the signal of interest has been removed, only whether significant unwanted variation remains; the plot cannot tell if the baby has been thrown out with the bath water. For example, the procedure of simply removing additive and non-additive effects from the gender data clearly removed a large amount of unwanted variation, as evidenced by the series of RLE plots, but we most likely also removed much of the signal of interest. Removing unwanted variation without also removing the signal of interest is a more sophisticated enterprise (for one approach, see \citealp{gagnon2012}). Thus, ``bad" looking RLE plots are usually \emph{strong} evidence that a normalisation procedure has failed, but ``good" looking RLE plots are only \emph{weak} evidence that a procedure has succeed, in the sense of not also removing signal of interest. Strong evidence that a procedure has succeeded needs to be obtained from other sources, e.g.\ $p$-value histograms, positive control gene rankings, comparison with previous results, and, best of all, experimental validation (see \citealp{gagnon2012}). 

Lastly, we mention two important points about assumption (A), i.e.\ that expression of a majority of genes are unaffected by the biological factors of interest. Firstly, this assumption is not always needed to infer the presence of unwanted variation from an RLE plot. Large differences between \emph{replicate} samples are an immediate sign of unwanted variation. In the gender study, for example, samples 5 and 26 are from the same individual and brain region; nominally, the samples are identical. The large disparity between the two can only be explained by unwanted variation, most likely resulting from being analysed at different laboratories. 

Secondly, assumption (A) is not always \emph{safe} to make. There may be instances where different biological factors of interest elicit a shift in the expression levels of a large majority of genes. In that case, a bad RLE plot might not be a reliable sign of unwanted variation, but a sign of a genuine shift in expression levels for some of the samples. Thankfully, however, these instances are rare, and when they do occur the effect is often expected.

\section{Conclusion}

We have seen that RLE plots, i.e.\ boxplots of deviations from gene medians, provide a simple, yet powerful, tool for detecting and visualising unwanted variation in high dimensional microarray data, the presence of which is often problematic. The only assumption we need to interpret sample heterogeneity in an RLE plot as a sign of unwanted variation is that expression levels of a majority of genes are unaffected by the biological factors of interest. We noted, however, that while this assumption is often plausible, it is sometimes not safe to make, and sometimes not even needed. We have seen that RLE plots can reveal unwanted variation in two ways, i.e.\ varying boxplot position and varying boxplot width, and that additive and non-additive sample effects can produce these features, although additive effects alone cannot produce variation in boxplot width. We showed how simulated data with these effects produce these features, and that these effects provide an explanation of these features for real data. We have emphasised that due to their ability to detect unwanted variation, RLE plots are particularly useful for assessing whether a normalisation procedure has been successful, with a ``bad" plot usually suggesting that the procedure has failed. However, we cautioned that while bad looking RLE plots are excellent evidence that a normalisation procedure has failed, good looking RLE plots are only weak evidence that a procedure has succeeded, in the sense of not also removing signal of interest. Although our discussion has been framed in terms of microarray expression data, the original context in which RLE plots were devised, we hope we have conveyed how RLE plots might be useful for revealing unwanted variation in many other kinds of high dimensional data, where such variation can be problematic.

\bibliographystyle{apalike}
\bibliography{references}

\end{document}